\documentclass[aps,prd,preprintnumbers,notitlepage,superscriptaddress,nofootinbib,longbibliography]{revtex4-1}

\usepackage{graphicx}
\usepackage[caption=false]{subfig}
\usepackage{amssymb}
\usepackage{amsfonts}
\usepackage{amsmath} 
\usepackage{color}
\usepackage[colorlinks=true,linkcolor=magenta,citecolor=cyan]{hyperref}
\usepackage{ascmac}
\usepackage[utf8]{inputenc}

\newcommand*{\alphad}{{\dot \alpha}}
\newcommand*{\alphadd}{{\ddot \alpha}}

\newcommand*{\phid}{{\dot \phi}}

\newcommand*{\vd}{{\dot v}}

\newcommand*{\Hd}{{\dot H}}

\newcommand*{\sigmad}{{\dot \sigma}}
\newcommand*{\sigmadd}{{\ddot \sigma}}

\newcommand*{\Mpl}{{M_{\rm pl}}}

\newcommand*{\sgn}{\,{\rm sgn}}

\begin{document}

\title{Accelerating Universe with a stable extra dimension in cuscuton gravity}

\author{Asuka Ito}
\email[Email: ]{asuka-ito@stu.kobe-u.ac.jp}
\affiliation{Department of Physics, Kobe University, Kobe 657-8501, Japan}

\author{Yuki Sakakihara}
\email[Email: ]{sakakihara@port.kobe-u.ac.jp}
\affiliation{Department of Physics, Kobe University, Kobe 657-8501, Japan}
\affiliation{Department of Mathematics and Physics, Osaka City University, Osaka 558-8585, Japan}

\author{Jiro Soda}
\email[Email: ]{jiro@phys.sci.kobe-u.ac.jp}
\affiliation{Department of Physics, Kobe University, Kobe 657-8501, Japan}

\preprint{KOBE-COSMO-09-11}

\begin{abstract}
We study Kaluza-Klein cosmology in cuscuton gravity and
 find an exact solution describing an accelerating
 4-dimensional universe with a stable extra dimension.
A cuscuton which is a non-dynamical scalar field is responsible for the accelerating expansion 
and a vector field makes the extra dimensional space stable.
Remarkably, the accelerating universe in our model is not exactly de Sitter.
\end{abstract}

\maketitle

\tableofcontents

%%%%%%%%%%%%%%%%%%%%%%%%%%%%%%%%%%%%%%%%%%%%%%%%%%%%%
\section{Introduction}
\label{Sec:1}
%%%%%%%%%%%%%%%%%%%%%%%%%%%%%%%%%%%%%%%%%%%%%%%%%%%%%
%
As is well known, superstring theory predicts our spacetime has 10-dimensions.
However, the idea of higher dimensional spacetime itself  is not new, in fact, goes back to
 Kaluza-Klein theory~\cite{Kaluza:1921tu,Klein:1926tv} which unifies 
electromagnetic force and gravity in 4-dimensional spacetime as pure gravity in 5-dimensional spacetime. 
Of course, since our real world has 4-dimensions, the extra dimensions must be 
invisible. This can be realized if the extra dimensions are compacified into
a small size. A natural mechanism for compactification of
the extra dimensions has been proposed in the context of cosmology. 
In the original proposal~\cite{Chodos:1979vk}, the 4-dimensional universe is expanding and the extra dimensions are contracting. 
Since then, Kaluza-Klein cosmology has been intensively investigated~\cite{Freund:1980xh,Maeda:1983fq,Maeda:1984un,Ishihara:1984wx}.
Now, the main issue is how to stabilize the contracting extra dimensions.
Actually, more natural scenario is as follows. 
Initially, all of spatial dimensions were compact and small.
Subsequently, in the course of cosmological evolution, only 4-dimensional universe has expanded up to the present scale. In this paper, we call this particular scenario ``Kaluza-Klein scenario''.
To our best knowledge, there seems no concrete model 
to realize this Kaluza-Klein scenario. 
In particular, 
it is difficult to construct 4-dimensional inflationary universe with stable extra dimensions. 
Indeed, when we put cosmological constant in higher dimensions,
the extra dimensions can be easily decompactified.
Recently, the mechanism for compactification has been also discussed in
the context of string theory and then
most discussions rely on the 4-dimensional effective action method.
First, the stability of extra dimensions is 
realized using a 4-dimensional non-perturbative mechanism.    
Next,  matters are considered in 4-dimensions to realize inflation.
However, the higher dimensional picture of this stabilization procedure is not obvious. 
Thus, it is still worth seeking the Kaluza-Klein scenario.

For resolving difficulties in cosmology such as the dark energy problem, 
modified theories of gravity have been extensively utilized.
We can expect that modified gravity also plays a role for realizing the Kaluza-Klein scenario.
Indeed, a compactification mechanism in Einstein-aether gravity~\cite{Jacobson:2000xp} has been proposed~\cite{Carroll:2008pk}. The aether field 
defines a preferred spacelike direction and
violates the rotational invariance of the 5-dimensional space.
It is shown that an attractive force of the aether field can stabilize the extra dimension. 
Unfortunately, the above model suffers from instabilities due to ghosts or
tachyons~\cite{Carroll:2009em,Himmetoglu:2008hx,Himmetoglu:2008zp,Himmetoglu:2009qi}.
Notice that, in 4-dimensions, 
this fact was known as the difficulty of constructing 
anisotropic inflation models (see review papers~\cite{Soda:2012zm,Maleknejad:2012fw}).
Remarkably, in 2009, there appeared a healthy model realizing anisotropic inflationary expansion 
with a vector field which induces a preferred direction due to a gauge kinetic function~\cite{Watanabe:2009ct}. 
Hence, it is legitimate to apply the anisotropic inflation model in modified gravity
to the Kaluza-Klein scenario. 
As to the modified gravity, in this paper, we focus on cuscuton gravity containing a non-dynamical scalar field, the so-called cuscuton~\cite{Afshordi:2006ad}.

The cuscuton gravity is a minimally modified theory of gravity in the sense that there are only 
two physical degrees of freedom of a tensor field.
It belongs to a subclass of scalar-tensor theories where the Lorentz invariance is 
broken~\cite{Afshordi:2009tt,Bhattacharyya:2016mah,Gomes:2017tzd,Mukohyama:2019unx} and 
can be extended into more general class~\cite{Iyonaga:2018vnu}.
Aspects of symmetry in cuscuton gravity is studied in~\cite{Pajer:2018egx}.
In particular, it has been shown that cuscuton gravity is useful in 
cosmology~\cite{Afshordi:2007yx,Boruah:2017tvg}, for instance, 
a healthy bouncing solution without instabilities were realized~\cite{Boruah:2018pvq}.
Moreover, any inflation models could be reconciled with observations by
virtue of cuscutons~\cite{Ito:2019fie}.
Hence, it is interesting to apply cuscuton gravity to the Kaluza-Klein scenario.

In our model of the Kaluza-Klein scenario, there is a vector field coupled with 
a cuscuton. 
We find the first exact solution describing an accelerating universe 
with a static extra dimension~\cite{Barrow:1986yt,Bringmann:2003sz}.
The vector field is responsible for the stability of the extra dimension
and the cuscuton drives the accelerating expansion of our 3-dimensional space.
Note that the universe is not a de Sitter spacetime
as a consequence of cuscuton gravity.
This is in contrast to the conventional compactification models.
Indeed, usually, we use 4-dimensional effective potentials for a radion,
the radius of the extra dimension, to
describe the dynamics of the extra dimensions during inflation
or the late time acceleration.
Then, the radion is constant only when 
it is at the minimum of the potential.
Consequently, the 4-dimensional spacetime have to be the de Sitter if the extra dimension is static.
Even if one puts additional matters into the effective theory, they will vanish eventually in light of 
the Wald's no-hair theorem~\cite{Wald:1983ky}.
It implies that we always obtain the 4-dimensional de Sitter spacetime 
if the extra dimension is static.
To avoid this consequence, one can tune the potential so that the minimum point is Minkowski.
In this case, we can add the 4-dimensional inflaton by hand to get a quasi-de Sitter universe.
However, this procedure has no clear meaning from the higher dimensional point of view.
The reason why we obtain non-conventional results can be attributed to the followings.
First, the presence of cuscutons can violate energy conditions 
assumed in the no-hair theorem~\cite{Wald:1983ky,Sakakihara:2012iq}.
Remarkably, even if the energy condition is violated, there is no instabilities in cuscuton gravity because
the cuscuton is non-dynamical~\cite{Boruah:2017tvg,Boruah:2018pvq,Ito:2019fie}.
Second, the cuscuton has a nontrivial coupling with the vector field, which 
also violate the assumption in the proof of no-hair theorem~\cite{Soda:2012zm,Maleknejad:2012fw}.

The paper is organized as follows:
In Sec.~\ref{sec2}, we explain our setup, the action including a cuscuton and a vector field in 
($4+1$)-dimensional spacetime, and derive the equations of motion. We then find exact power-law solutions. 
In Sec.~\ref{sec3}, we give accelerating universe solutions with a static extra space dimension. 
In Sec.~\ref{sec4}, we investigate the stability of the solutions we found in Sec.~\ref{sec3}.
It turns out that the solutions are attractors in phase space. 
Sec.~V is devoted to the conclusion. 
The discussion is extended into general $(n+1)$-dimensional spacetime in Appendix.~\ref{App:1}.

%%%%%%%%%%%%%%%%%%%%%%%%%%%%%%%%%%%%%%%%%%%%%%%%%%%%%
\section{Anisotropic ($4+1$)-dimensional spacetime with cuscuton} \label{sec2}
%%%%%%%%%%%%%%%%%%%%%%%%%%%%%%%%%%%%%%%%%%%%%%%%%%%%%
%
For simplicity, we start from ($4+1$)-dimensional spacetime with the action including a cuscuton. 
The case of general $(n+1)$-dimensional spacetime will be studied in Appendix~\ref{App:1}.

In order to violate rotational invariance
and realize the compactification, we also include a vector field coupled with the cuscuton. Then, we have 
\begin{align}
 \mathcal{S}=\int d^5x \sqrt{-g} \Biggl[\frac{1}{2\kappa}R\pm \mu^2\sqrt{-g^{\mu\nu}\partial_\mu \phi\partial_\nu \phi}-V(\phi) -\frac{1}{4}f(\phi) F_{\mu\nu}F^{\mu\nu}\Biggr]  \ , \label{ac}
\end{align}
where we have defined $\kappa=1/\Mpl_5^3$ using five dimensional Planck scale $M_{\rm pl5}$.
$\mu$ is a parameter associated with the kinetic term of the cuscuton, which has the mass dimension~$[\mu^2]=5/2$. 
The field strength of the $U(1)$ vector field~$A_\mu$ is defined by
\begin{align}
 F_{\mu\nu}=\partial_\mu A_\nu-\partial_\nu A_\mu \ . \label{F}
\end{align}
The cuscuton is coupled with the kinetic term of the vector field though a function~$f(\phi)$.
Note that the mass dimensions of the fields are $[\phi]=3/2$ and $[A_\mu]=3/2$.
We assume that homogeneous electric fields exist along with the extra space dimension
and the cuscuton is supposed to depend only on time,
\begin{align} 
 A_{\mu}=(0,0,0,0, v(t)) \ , \qquad   \phi=\phi(t) \ .  \label{A}
\end{align}
With these ansatzes, 
we take a homogeneous but anisotropic metric ansatz discriminating the extra dimension:
\begin{align}
 ds^2=-dt^2+e^{2\alpha(t)}\Bigl(e^{-6\sigma(t)}dx_4^2+e^{2\sigma(t)}(dx^2+dy^2+dz^2)\Bigr) \ .   \label{met}
\end{align}
We expect that the expansion of the extra dimension slows down due to the vector field.
Substituting (\ref{F})--(\ref{met}) into (\ref{ac}), we obtain the action,
\begin{align}
 \mathcal{S}=\int d^5x\; e^{4\alpha}\Biggl[\frac{1}{\kappa}(-6\alphad^2+6\sigmad^2)\pm\mu^2{\rm sgn}(\phid)\phid -V(\phi)+\frac{1}{2}f(\phi)e^{-2\alpha+6\sigma}\vd^2\Biggr]\ ,
\end{align}
and the equations of motion derived from this action are 
\begin{align}
 \alphadd&=-4\alphad^2 \mp \frac{\kappa\mu^2}{3}\sgn (\phid)\phid + \frac{2\kappa}{3}V(\phi)+\frac{\kappa}{12}f(\phi)e^{-2\alpha+6\sigma}\vd^2 \ ,\label{143947_5Jun19}\\
 \alphad^2&=\sigmad^2+\frac{\kappa}{6}\biggl[V(\phi)+\frac{1}{2}f(\phi)e^{-2\alpha+6\sigma}\vd^2\biggr]\ ,\label{084144_8Apr19}\\
 \sigmadd&=-4\alphad\sigmad+\frac{\kappa}{4}f e^{-2\alpha+6\sigma}\vd^2 \ ,\\
 &\pm 4 \mu^2 \alphad \; {\rm sgn} (\phid) + V_{\phi}-\frac{1}{2}f_{\phi}e^{-2\alpha+6\sigma}\vd^2=0 \ ,\label{144148_5Jun19}\\
 &(e^{2\alpha+6\sigma}f\vd)^\cdot=0 \ . \label{090339_8Apr19}
\end{align}
The second equation is the Friedmann equation in Einstein's gravity, which is not independent from the others.
We integrate \eqref{090339_8Apr19} and have 
\begin{align}
 \vd = f^{-1}e^{-2\alpha-6\sigma}C \ , 
\end{align}
where $C$ is a constant of integration with the mass dimension~$[C]=5/2$.
After substituting the solution into \eqref{143947_5Jun19}--\eqref{144148_5Jun19}, we obtain the four equations:
\begin{align}
 \alphadd&=-4\alphad^2 \mp \frac{\kappa\mu^2}{3}\sgn (\phid)\phid + \frac{2\kappa}{3}V+\frac{\kappa}{12}f^{-1}e^{-6\alpha-6\sigma}C^2 \ ,\label{100723_15Apr19}\\
 \alphad^2&=\sigmad^2+\frac{\kappa}{6}\biggl[V+\frac{1}{2}f^{-1}e^{-6\alpha-6\sigma}C^2\biggr]\ ,\label{100757_15Apr19}\\
 \sigmadd&=-4\alphad\sigmad+\frac{\kappa}{4}f^{-1} e^{-6\alpha-6\sigma}C^2 \ ,\label{100735_15Apr19}\\
  &\pm 4 \mu^2 \alphad \; {\rm sgn} (\phid) + V_{\phi}-\frac{1}{2}f_\phi f^{-2}e^{-6\alpha-6\sigma}C^2=0 \ . \label{102736_15Apr19}
\end{align}
From now on, to seek for exact power-law solutions of above equations,
we choose the potential of the cuscuton and the gauge kinetic function as 
\cite{Iyonaga:2018vnu,Kanno:2010nr,Ohashi:2013pca,Ito:2015sxj}
\begin{align}
 V=\frac{1}{2}m^2\phi^2\ ,\qquad 
 f=f_0\Bigl(\frac{\phi}{\Mpl_{5}^{3/2}}\Bigr)^{2w}  \ , \label{po}
\end{align} 
respectively, where $m$ is the mass of the cuscuton, $f_{0}$ is a positive constant and $w$ is an integer.
Then, we take the ansatz: 
 \begin{align}
 \alpha=p_1 \log \Mpl_5 t \ , \qquad 
 \sigma=p_2 \log \Mpl_5 t \ , \qquad
 \phi=\frac{q}{t}\Mpl_5^{1/2}  \ ,   \label{pa}
\end{align}
where $p_1 , p_2 , q$ are parameters. 
Eqs.\,\eqref{100723_15Apr19}--\eqref{102736_15Apr19} accommodate power-law solutions
with the ansatz  (\ref{pa}).
First of all, we need a relation,
\begin{align}
 &w -3p_1-3p_2 +1 =0 \ . \label{133718_18Apr19} 
\end{align}
We also find the following set of the algebraic equations relating the parameters:
\begin{align}
 -p_1+4p_1^2 &=-4\xi|q|+4\lambda |q|^2+\frac{\gamma}{|q|^{2w}} \ ,\label{144529_28May19} \\
 p_1^2-p_2^2&=\lambda |q|^2+\frac{\gamma}{|q|^{2w}} \ , \\
 -p_2+4p_1p_2&=3\frac{\gamma}{|q|^{2w}} \ ,\label{144452_28May19}\\
 4|q|\xi p_1 &=\lambda |q|^2 -w\frac{\gamma}{|q|^{2w}} \ ,\label{141656_16Apr19} 
\end{align}
where we have defined new parameters as
\begin{align}
 & \gamma=\frac{(C^2/\Mpl_5^{5})}{12 f_0 } \ , \qquad
 \lambda=\frac{(m^2/\Mpl_{5}^2)}{12} \ , \qquad 
 \xi= \pm \frac{(\mu^2/\Mpl_{5}^{\frac{5}{2}})}{12} \ .\label{205113_5Jun19}
\end{align} 
We have $7$ parameters in the equations: $\{p_1, p_2, |q|, \gamma, w , \lambda, \xi \}$ and $4$-independent equations~\eqref{133718_18Apr19}, \eqref{144529_28May19}, \eqref{144452_28May19} and \eqref{141656_16Apr19}. 
Let us solve the algebraic equations about $\{\gamma, w , \lambda, \xi \}$ and 
express them by $\{p_1, p_2, |q|\}$.
One can solve \eqref{133718_18Apr19} for $w$, and \eqref{144452_28May19} for $\gamma$,
\begin{align}
 w=3(p_1+p_2)-1 \ ,\label{134529_18Apr19} 
\end{align}
\begin{align}
 \gamma=\frac{1}{3}p_2(-1+4 p_1)|q|^{2w} \ ,\label{134522_18Apr19}
\end{align}
respectively. 
Furthermore, from \eqref{144529_28May19}, \eqref{141656_16Apr19}, 
(\ref{134529_18Apr19}) and (\ref{134522_18Apr19}), we obtain
\begin{align}
 \lambda=\frac{3(p_1^2-p_2^2)+p_2(1-4 p_1)}{3|q|^2} \ \label{205342_6Jun19}
\end{align}
and
\begin{align}
 \xi=\frac{(p_1+p_2)(1-4p_2)}{4|q|}\ .\label{205145_5Jun19}
\end{align}
Apparently, there exists power-law solutions for an arbitrary set of parameters~$\{p_1,p_2,|q|\}$.
However, the parameter region is restricted because of the positivity of $\gamma$ and $\lambda$.
In the next section, we focus on power-law solutions
corresponding to expanding four dimensional spacetime with a static extra dimension 
and reveal the allowed parameter region.

%%%%%%%%%%%%%%%%%%%%%%%%%%%%%%%%%%%%%%%%%%%%%%%%%%%%%
\section{Kaluza-Klein scenario in cuscuton gravity} \label{sec3}
%%%%%%%%%%%%%%%%%%%%%%%%%%%%%%%%%%%%%%%%%%%%%%%%%%%%%
%
In the previous section, we have obtained power-law solutions for the system with a cuscuton and a vector field. Depending on the parameters of the solution, $\{p_1,p_2,|q|\}$, various situations can be realized.
Hence, we are now in a position to realize the Kaluza-Klein scenario.
 
To achieve our aim,
we need to make the extra dimension represented by a coordinate $x_4$ frozen, 
while the other spatial dimensions are expanding as in our universe. 
Such situation occurs for $p_1=3 p_2$ in the power-law solutions.
In that case, the spacetime metric is described by
\begin{align}
 ds^2=-dt^2+dx_4^2+( \Mpl_5 t)^{8 p_2}(dx^2+dy^2+dz^2)\ .\label{084343_6Jun19}
\end{align}
The Hubble parameter in the four dimensional spacetime is 
\begin{align}
 H_4=\frac{4p_2}{t} \ ,\label{145542_10Jun19}
\end{align}
which should be positive to describe the expanding universe, i.e., $p_{2} > 0$.
Under the assumption, $p_1=3p_2$, the relations between the parameters~\eqref{134529_18Apr19}--\eqref{205145_5Jun19}
are reduced to  
 \begin{align}
  w=-1+12 p_2 \ , \qquad 
  \gamma=\frac{p_2 [-1+12 p_2]}{3}|q|^{2(-1+12p_2)} \ , \qquad  
  \lambda=\frac{p_2 [1+12 p_2]}{3|q|^2} \ , \qquad  
  \xi=\frac{p_2(1-4 p_2)}{|q|} \ .\label{145115_10Jun19}
 \end{align}
Since the parameters~$\gamma$ and $\lambda$ should be positive by definition, 
$p_{2}$ should satisfies
\footnote{ 
\begin{align}
 p_2<-\frac{1}{12} \nonumber  
\end{align}
is also another possibility, but this case realizes the contraction of $3$-dimensional
space coordinate, and therefore, 
it is out of our interest here.}
 \begin{align}
  p_2>\frac{1}{12} \ . \label{con}
 \end{align}
It implies $w$ is always positive.

If we divide this regime depending on the sign of $\xi$, they would be 
   \begin{align}
    \frac{1}{12}<p_2<\frac{1}{4} \quad ({\rm for }\;\xi> 0)\ , \qquad     
    \frac{1}{4}\leq p_2 \quad ({\rm for} \;\xi\leq 0)\ . \label{145822_10Jun19}
   \end{align}
Since the second time derivative of the scale factor is proportional to $p_{2}^{2}(4p_2-1)$, we find that the the existence of a cuscuton with $\xi<0$ is essential to realize the accelerated expansion of the four dimensional spacetime. 
Note that there is an interesting case $\xi = 0$ $(p_{2} = 1/4)$, i.e., $\mu^{2} = 0$, where  
the kinetic term of the cuscuton vanishes.
In this case, after integrating out $\phi$ in the action, 
the theory only includes a vector field with non-linear terms of $F_{\mu\nu}F^{\mu\nu}$.

The solution (\ref{145115_10Jun19}) exists 
in the two dimensional parameter space $\{ p_{2}, |q| \}$ satisfying the condition (\ref{con}).
Here we mention that one can always characterize the solution by using 
$\{ w, \lambda \}$ which appear in the original action (\ref{ac}) in contrast
to $\{ p_{2}, |q| \}$.
Actually, in terms of $\{ w, \lambda \}$, the solution is written by~\footnote{
For the set of model parameters~$(w, \lambda)$, we cannot freely choose the value of $\xi$ to realize the Kaluza-Klein scenario. Even if the parameter~$\xi$ does not exactly satisfy $\xi^2=\xi^2(w,\lambda)$, where $\xi^2(w, \lambda)$ is defined by \eqref{204845_6Jun19}, we can approximately realize the Kaluza-Klein scenario as far as $\xi^2\sim \xi^2(w, \lambda)$. Then, the ratio of the expansion rate of the extra space dimension to that of our $3$-dimensional space, $(p_1-3p_2)/(p_1+p_2)$, is small enough. 
}
\begin{align}
 p_2(w)=\frac{1+w}{12}\ , \quad 
 \xi^2=\frac{3p_2(w)(1-4p_2(w))^2}{(1+12p_2(w))}\lambda \ , \quad
 |q|=\frac{p_2(w)(1-4p_2(w))}{\xi(w,\lambda)} \ .  \label{204845_6Jun19}
\end{align}

It is useful to evaluate the slow roll parameters, which characterize inflationary universe, 
and the ratio of the energy density of the vector field to that of the cuscuton.
The slow roll parameters of the four dimensional spacetime is calculated from~\eqref{145542_10Jun19} as
\begin{align}
 \epsilon_4=-\frac{\Hd_4}{H_4^2}=\frac{1}{4 p_2} = \frac{3}{ 1+w } \ , \quad 
 \eta=2\epsilon_4 -\frac{\dot{\epsilon_4}}{2H_4\epsilon_4}=2\epsilon_4 \ ,
\end{align}
where we have used the fact that $\dot{\epsilon_4}=0$ for power-law solutions.
%We see that the four dimensional spacetime is expanding with an accelerating rate determined by $p_{2}$, 
%while the extra dimension is completely static.
In particular, inflationary universe is realized if 
$4 p_2\gg 1$ is satisfied, which implies $\xi \ll 0 \Leftrightarrow w \gg 2$.
%Then from (\ref{enecon}), the Null energy condition is violated and thus the solution is quite non-trivial.
Finally, The ratio of the energy density of the vector field to that of the cuscuton is found to be
\begin{align}
 \frac{\rho_A}{\rho_c}=\frac{\gamma}{\lambda |q|^{2(1+w)}}=\frac{-1+12 p_2}{1+12 p_2} = \frac{w}{2+w} \ .
\end{align}
It is almost the unity in the inflationary universe, $p_{2} \gg 1$ ($w \gg 1$).

In the next section, we investigate the stability of the solution (\ref{145115_10Jun19}).
We will see that the solution is an attractor in phase space as long as the inequality (\ref{con}) is satisfied.
%%%%%%%%%%%%%%%%%%%%%%%%%%%%%%%%%%%%%%%%%%%%%%%%%%%%%
\section{Stability of the solution}   \label{sec4}
%%%%%%%%%%%%%%%%%%%%%%%%%%%%%%%%%%%%%%%%%%%%%%%%%%%%%
%
In the previous section, 
we found exact power-law solutions describing expanding universe with a static extra dimension.
Let us examine if the solutions are stable or not.

It is convenient to recast the equations (\ref{100723_15Apr19})--(\ref{102736_15Apr19}) 
with following new dimensionless variables:
\begin{align}
 X=\frac{\sigmad}{\alphad} \ , \qquad  
 Y =  - {\rm sgn}(\dot{\phi}) \frac{\phi}{\alphad \Mpl_5^{1/2}} \ , \qquad 
 \tilde{Y}=  {\rm sgn}(\dot{\phi})  \frac{\dot{\phi}}{\dot{\alpha}^{2} \Mpl_5^{1/2}} \ , \qquad 
 Z=\frac{C^2}{12\Mpl_5^5}\frac{\Mpl^2}{\alphad^2}f^{-1}e^{-6\alpha-6\sigma} \ .
\end{align}
Using these variables, Eqs.(\ref{100723_15Apr19})--(\ref{102736_15Apr19}) can be rewritten as 
\begin{eqnarray}
  \frac{dX}{d\alpha} &=& -4X + 3Z -X 
    \left[ -4  \xi \tilde{Y}
           + 4 \lambda Y^{2} + Z  -4 \right] \ ,  \label{e1} \\
  1 &=& X^{2} + \lambda  Y^{2} + Z     \ ,  \label{c1}    \\
  \frac{dZ}{d\alpha} &=& 2 Z 
    \left[ 1 + 4  \xi  \tilde{Y}
           -  4 \lambda Y^{2} - Z 
           + w \frac{\tilde{Y}}{Y}  
           - 3 X     \right]                            \ ,  \label{e2} \\
  0 &=& - 4  \xi 
          + \lambda Y  -  \frac{w Z}{Y}    \ ,   \label{c2}
\end{eqnarray}
where we have used the e-folding number $\alpha$ as the time coordinate of the system.
There are two constraint equations, one is the Friedmann equation (\ref{c1}) and 
the other is the equation of motion of the cuscuton (\ref{c2}).
Note that the cuscuton is non-dynamical scalar field and thus (\ref{c2}) is just a constraint equation.
From (\ref{c2}), we can express $Y$ in terms of $Z$ as
\begin{equation}
  Y = \frac{
     2  \xi
    + \sqrt{ w\lambda Z + 4\xi^{2} } }
    {\lambda}   \ , \label{phda}
\end{equation}
where we have used the fact $Y > 0$.
Moreover, differentiating (\ref{phda}) with respect to $t$, one can reduce $\tilde{Y}$ as a 
function of $X$ and $Z$:
\begin{equation}
 \tilde{Y} =
\frac {  Y 
\left[  \left( 4 \lambda Y^{3} +  YZ  -4 Y \right) \sqrt{ w\lambda Z + 4\xi^{2} }
-4 \lambda w Y^{2} Z
-3 w X Z 
-w Z^{2} 
+ w Z
 \right] 
 }
 {
  \left( 4 \xi Y^2  -Y  \right) \sqrt{ w\lambda Z + 4 \xi^{2} }
-4  \xi w Y Z
-  w^{2}Z  } 
   \ . \label{phda2}
\end{equation}
Finally, substituting (\ref{phda}) and (\ref{phda2}), 
into (\ref{e1}), (\ref{c1}) and (\ref{e2}),  we get a closed autonomous system with a constraint equation 
relating $X$ with $Z$.

On the other hand,
the power-law solution (\ref{145115_10Jun19}) can be expressed in terms of $X$ and $Z$ as
\begin{align}
 & X  = \frac{1}{3} \ , \qquad 
 Z  = \frac{w}{27 p_2(w)}   \ .  \label{48}
\end{align}
It is easy to check that 
the power-law solution (\ref{48}) is indeed a fixed point in the phase space, which is 
defined by $\frac{dX}{d\alpha} = \frac{dZ}{d\alpha} = 0$.

In order to examine the stability of the fixed point,
we first eliminate $X$ from (\ref{e2}) by using (\ref{c1}), (\ref{phda}) and (\ref{phda2}).
Expanding $Z$ around the power-law solution (\ref{48}) as $Z=w/27p_{2}+\delta Z$,  we have
\begin{align}
 \frac{d \delta Z}{d\alpha}=-\frac{[-1+12 p_2]}{3p_2} \delta Z = - \frac{w}{3p_2} \delta Z \ , \label{49}
\end{align}
at the linear order. 
(\ref{49}) shows that the power-law solution is stable if the inequality $p_{2} > 1/12$ is satisfied.
The condition coincides with the existence condition of the power-law solution (\ref{con}), so that 
the stability is guaranteed for the expanding power-law solutions~\eqref{145115_10Jun19} with \eqref{con}.
We have also numerically confirmed that the solution (\ref{48}) is attractor in the phase space.
It proved the stability at the nonlinear level.

%%%%%%%%%%%%%%%%%%%%%%%%%%%%%%%%%%%%%%%%%%%%%%%%%%%%%
\section{Conclusion}
\label{Sec:conc}
%%%%%%%%%%%%%%%%%%%%%%%%%%%%%%%%%%%%%%%%%%%%%%%%%%%%%
%
We studied the Kaluza-Klein scenario in cuscuton gravity.
In our model of Kaluza-Klein scenario, 
a vector field coupled with a cuscuton has a vacuum expectation value 
along with the direction of the extra dimension and violate the rotational invariance of 
higher dimensional spaces.
We found an exact power-law solution of the Einstein and the field equations.
It was shown that the solution describes accelerating expansion of four dimensional spacetime with a completely static extra dimension.
To the best of our knowledge, this is the first concrete model
of accelerating universes other than the de Sitter spacetime
with a static extra dimension~\cite{Barrow:1986yt,Bringmann:2003sz}.

The stability of the solution was also investigated in Sec.~\ref{sec4}.
We performed dynamical system analysis~(\ref{100723_15Apr19})--(\ref{102736_15Apr19})
and revealed the condition for stability~(\ref{49}).
It coincides with the condition for existence of the solution~(\ref{con}).
Therefore, the solution is always stable if exists.
All the discussion in the text focused on ($4+1$)-dimensional spacetime for simplicity, however,
it is easy to extend it to general ($n+1$)-dimensional spacetime (see App.~\ref{App:1}).

Although we used a $1$-form field to compactify one extra dimension in this paper,
one can apply $p$ $1$-form fields~\cite{Yamamoto:2012tq} or 
a $p$-form field~\cite{Ohashi:2013mka} to compactify $p$ extra dimensions simultaneously.
It is interesting to explore cosmological perturbations in the Kaluza-Klein scenario.
We leave these issues for future work.

\acknowledgements
A.\,I. was supported by Grant-in-Aid for JSPS Research Fellow and JSPS KAKENHI Grant No.JP17J00216.
Y.S. was supported by JSPS KAKENHI Grant Number JP17H02894.
 J.S. was in part supported by JSPS KAKENHI Grant Numbers JP17H02894, JP17K18778, JP15H05895, JP17H06359, JP18H04589. J.S. is also supported by JSPS Bilateral Joint Research Projects (JSPS-NRF col- laboration) String Axion Cosmology.

\appendix

%%%%%%%%%%%%%%%%%%%%%%%%%%%%%%%%%%%%%%%%%%%%%%%%%%%%%
\section{$(n+1)$-dimensional spacetime}
\label{App:1}
%%%%%%%%%%%%%%%%%%%%%%%%%%%%%%%%%%%%%%%%%%%%%%%%%%%%%

Though we have started with ($4+1$)-dimensional spacetime in the text, the discussion can be extended to the 
an arbitrary number of space dimensions. 
The argument is almost the same, and thus we shortly summarize the results.

The action in $(n+1)$-dimensional spacetime reads $(n>2)$
\begin{align}
 \mathcal{S}=\int d^{n+1}x \sqrt{-g} \Biggl[\frac{1}{2\kappa_n}R\pm \mu^2\sqrt{-g^{\mu\nu}\partial_\mu \phi\partial_\nu \phi}-V(\phi) -\frac{1}{4}f(\phi) F_{\mu\nu}F^{\mu\nu}\Biggr]  \ ,
\end{align}
where $\kappa_n=1/\Mpl_{n+1}^{n-1}$, $\Mpl_{n+1}$ is $(n+1)$-dimensional Planck scale, and the definition of field strength of the vector field~$A_\mu$ is just the higher dimensional counterpart. We note that the mass dimension of the field are $[\phi]=(n-1)/2$, $[A_\mu]=(n-1)/2$, and the parameter~$\mu$ has the mass dimension~$[\mu^2]=(n+1)/2$.  The spacetime metric is defined with one anisotropic direction,
\begin{align}
 ds^2=-dt^2+e^{2\alpha(t)}\Bigl(e^{-2(n-1)\sigma(t)}dx_n^2+e^{2\sigma(t)}(dx_1^2+\cdots+dx_{n-1}^2)\Bigr) \ ,
\end{align}
As the case of ($4+1$)-dimensional spacetime,
we take  the configuration of the vector field and the cuscuton as,
\begin{align}
 A_{\mu}=(0,\cdots,0 , v(t)) \ , \qquad \phi=\phi(t) \ ,
\end{align}
Then, the background action is written as
\begin{align}
 \mathcal{S}=\int d^{n+1}x\; e^{n\alpha}\Biggl[\frac{n(n-1)}{2\kappa_n}(-\alphad^2+\sigmad^2)\pm\mu^2{\rm sgn}(\phid)\phid -V(\phi)+\frac{1}{2}f(\phi)e^{-2\alpha+2(n-1)\sigma}\vd^2\Biggr]\ ,
\end{align}
which leads to the equations of motion as follows:
\begin{align}
 \alphadd&=-n\alphad^2 \mp \frac{\kappa_n\mu^2}{n-1}\sgn (\phid)\phid + \frac{2\kappa_n}{n-1}V(\phi)+\frac{\kappa_n}{n(n-1)}f(\phi)e^{-2\alpha+2(n-1)\sigma}\vd^2 \ ,\label{151822_10Jun19}\\
 \alphad^2&=\sigmad^2+\frac{2\kappa_n}{n(n-1)}\biggl[V(\phi)+\frac{1}{2}f(\phi)e^{-2\alpha+2(n-1)\sigma}\vd^2\biggr]\ ,\\
 \sigmadd&=-n\alphad\sigmad+\frac{\kappa_n}{n}f e^{-2\alpha+2(n-1)\sigma}\vd^2 \ ,\\
 &\pm n \mu^2 \alphad \; {\rm sgn} (\phid) + V_{\phi}-\frac{1}{2}f_{\phi}e^{-2\alpha+2(n-1)\sigma}\vd^2=0 \ ,\label{102558_15Apr19} \\
 &(e^{(n-2)\alpha+2(n-1)\sigma}f\vd)^\cdot=0 \ .\label{144952_28May19}
\end{align}
We integrate \eqref{144952_28May19} and have 
\begin{align}
 \vd = f^{-1}e^{-(n-2)\alpha-2(n-1)\sigma}C \ , 
\end{align}
where $C$ is a constant of integration with the mass dimension~$[C]=(n+1)/2$.
After substituting this solution into (\ref{151822_10Jun19})--(\ref{102558_15Apr19}), 
we list the equations:
\begin{align}
 \alphadd&=-n\alphad^2 \mp \frac{\kappa_n\mu^2}{n-1}\sgn (\phid)\phid + \frac{2\kappa_n}{n-1}V(\phi)+\frac{\kappa_n}{n(n-1)}f^{-1}e^{-2(n-1)\alpha-2(n-1)\sigma}C^2 \ ,\\
 \alphad^2&=\sigmad^2+\frac{2\kappa_n}{n(n-1)}\biggl[V(\phi)+\frac{1}{2}f^{-1}e^{-2(n-1)\alpha-2(n-1)\sigma}C^2\biggr]\ ,\\
  \sigmadd&=-n\alphad\sigmad+\frac{\kappa_n}{n}f^{-1} e^{-2(n-1)\alpha-2(n-1)\sigma}C^2 \ ,\\
  &\pm n \mu^2 \alphad \; {\rm sgn} (\phid) + V_{\phi}-\frac{1}{2}f_{\phi}f^{-2}e^{-2(n-1)\alpha-2(n-1)\sigma}C^2=0 \ .
\end{align}
The potential of the cuscuton and the function in front of the kinetic term of the vector field are defined by
\begin{align}
 V=\frac{1}{2}m^2\phi^2 \ ,\qquad
 f=f_0\Bigl(\frac{\phi}{\Mpl_{n+1}^{\frac{n-1}{2}}}\Bigr)^{2w}  \ .
\end{align} 

%%%%%%%%%%%%%%%%%%%%%%%%%%%%%%%%%%%%%%%%%%%%%%%%%%%%%
\subsection{Power-law solutions}
%\label{Sec:3}
%%%%%%%%%%%%%%%%%%%%%%%%%%%%%%%%%%%%%%%%%%%%%%%%%%%%

We search for power-law solutions with 
 \begin{align}
 \alpha&=p_1 \log \Mpl_{n+1} t \ , \\
 \sigma&=p_2 \log \Mpl_{n+1} t \ , \\
 \phi&=\frac{q}{t}\Mpl_{n+1}^{\frac{n-3}{2}}  \ ,
\end{align}
and the potential of the cuscuton~$V$ and the function~$f$ are reduced to 
\begin{align}
 V=\frac{1}{t^2}\frac{m^2 q^2 \Mpl_{n+1}^{n-3}}{2} \ , \qquad 
 f =f_0q^{2w} (\Mpl_{n+1} t)^{-2w}  \ .
\end{align} 
To have power-law solutions, we should require
\begin{align}
 & w -(n-1)p_1-(n-1)p_2 +1 =0 \ ,\label{100050_3Jun19}
\end{align}
and then, we have
\begin{align}
 -p_1+np_1^2&= -n\xi|q| +n\lambda q^2 +\frac{\gamma}{q^{2w}} \ ,\label{100121_3Jun19} \\
 p_1^2-p_2^2&=\lambda q^2+ \frac{\gamma}{q^{2w}}\ , \\
 -p_2+np_1p_2&=(n-1)\frac{\gamma}{q^{2w}} \ ,\label{100102_3Jun19}\\
 n \xi|q| p_1 &=\lambda q^2 -w\frac{\gamma}{q^{2w}} \ ,\label{100126_3Jun19}
\end{align}
where
\begin{align}
 &\gamma=\frac{(C^2/\Mpl_{n+1}^{n+1})}{n(n-1) f_0} \ , \qquad
 \lambda=\frac{(m^2/\Mpl_{n+1}^2)}{n(n-1)} \ , \qquad 
 \xi= \pm \frac{(\mu^2/\Mpl_{n+1}^{\frac{n+1}{2}})}{n(n-1)} \ .\label{100228_3Jun19}
\end{align} 
We have $7$ parameters: $\{p_1, p_2, q , \gamma, w , \lambda, \xi \}$ and $4$-independent equations~\eqref{100050_3Jun19}, \eqref{100121_3Jun19}, \eqref{100102_3Jun19}, and \eqref{100126_3Jun19}. 
We solve \eqref{100050_3Jun19} for $w$ and \eqref{100102_3Jun19} for $\gamma$, 
\begin{align}
 w=(n-1)(p_1+p_2)-1 \ ,\label{100222_3Jun19} 
\end{align}
\begin{align}
 \gamma=\frac{1}{n-1}p_2(-1+n p_1)q^{2w} \ , \label{gam}
\end{align}
respectively. Using \eqref{100222_3Jun19} and \eqref{gam} in \eqref{100121_3Jun19} and \eqref{100126_3Jun19},
we obtain
\begin{align}
 \lambda=\frac{(n-1)(p_1^2-p_2^2)+p_2(1-n p_1)}{(n-1)|q|^2} \ ,
\end{align}
\begin{align}
 \xi=\frac{(p_1+p_2)(1-np_2)}{n |q|}\ .
\end{align}
%%%%%%%%%%%%%%%%%%%%%%%%%%%%%%%%%%%%%%%%%%%%%%%%%%%%%
\subsection{Kaluza-Klein solutions}
%%%%%%%%%%%%%%%%%%%%%%%%%%%%%%%%%%%%%%%%%%%%%%%%%%%%%
We seek for the solution respecting $p_1=(n-1)p_2$, which implies the spacetime metric is written as
\begin{align}
 ds^2=-dt^2+dx_n^2+(\Mpl_{n+1} t)^{2 n p_2}(dx_1^2+dx_2^2+\cdots +dx_{n-1}^2)\ .
\end{align}
The effective Hubble parameter in the $n$-dimensional spacetime is 
\begin{align}
 H_{n}=n p_2 \ .
\end{align}
Then, the relations between the parameters becomes
 \begin{align}
  &w=-1+n(n-1) p_2 \ , \qquad 
  \gamma=\frac{p_2 [-1+n(n-1) p_2]}{n-1}|q|^{2(-1+n(n-1)p_2)} \ , \nonumber\\
  &\lambda=\frac{p_2 [1+n(n-1)(n-3) p_2]}{(n-1)|q|^2} \ , \qquad  
  \xi=\frac{p_2(1-n p_2)}{|q|} \ .\label{085610_6Jun19}
 \end{align}
From these relation, we can read off the condition for the existence of the solution from the 
positivity of $\gamma$ and $\lambda$ as
 \begin{align}
  p_2>\frac{1}{n(n-1)} \ ,
 \end{align}
which implies $w$ is always positive.
If we divide this condition depending on the sign of $\xi$, they would be 
   \begin{align}
    \frac{1}{n(n-1)}<p_2<\frac{1}{n} \quad ({\rm for }\;\xi> 0)\ , \qquad     
    \frac{1}{n}\leq p_2 \quad ({\rm for} \;\xi\leq 0)\ .\label{091037_6Jun19}
   \end{align}
The slow roll parameter in $n$-dimensional spacetime is 
\begin{align}
 \epsilon_{n}=\frac{1}{n p_2}=\frac{n-1}{1+w} \ ,
\end{align}
which means that the slow roll condition is $n p_2\gg 1$. 
From \eqref{085610_6Jun19}, we find that we need to satisfy $\xi<0 \Leftrightarrow w>n-2$ for the realization of the accelerated expansion in $n$-dimensional spacetime.
%%%%%%%%%%%%%%%%%%%%%%%%%%%%%%%%%%%%%%%%%%%%%%%%%%%%%
\subsection{Stability}
%%%%%%%%%%%%%%%%%%%%%%%%%%%%%%%%%%%%%%%%%%%%%%%%%%%%%
%
Even in $(n+1)$-dimensional spacetime, the stability of the power-law solution~(\ref{085610_6Jun19})
does not change at all, since the perturbative equation is almost the same as (\ref{49}):
\begin{align}
 \frac{d \delta Z}{d\alpha}=-\frac{[-1+n(n-1)p_2]}{(n-1)p_2} \delta Z = - \frac{w}{(n-1)p_2} \delta Z \ ,
\end{align}
where $\delta Z=Z-\bar{Z}$, which is the perturbation of 
\begin{equation}
  Z=\frac{C^2}{n(n-1)\Mpl_{n+1}^{n+1}}\frac{\Mpl^2}{\alphad^2}f^{-1}e^{-2(n-1)\alpha-2(n-1)\sigma} \ ,
\end{equation}
around $\bar{Z}=w/[(n-1)^3 p_2(w)]$, where $p_2(w)=(1+w)/[n(n-1)]$, defined by the power-law solution (\ref{085610_6Jun19}).

\nocite{*}

\bibliography{Cuscuton_bib}

\end{document}